\documentclass[twocolumn,showpacs,prl]{revtex4}

\usepackage{graphicx}
\usepackage{dcolumn}
\usepackage{amsmath}
\usepackage{amssymb}
\usepackage{epsfig}

\begin{document}

\title{Aeolian sans ripples: experimental study of saturated states}

\author{Bruno Andreotti}
\author{Philippe Claudin}
\affiliation{Laboratoire de Physique et M\'ecanique des Milieux H\'et\'erog\`enes (UMR CNRS 7636), ESPCI, 10 rue Vauquelin 75231 Paris Cedex 05, France.}
\author{Olivier Pouliquen}
\affiliation{IUSTI, Universit\'e de Provence -- CNRS,
5 rue Enrico Fermi, 13453 Marseille Cedex 13, France.}

\date{\today}

\begin{abstract}
We report an experimental investigation of aeolian sand ripples, performed both in a wind tunnel and on stoss slopes of dunes. Starting from a flat bed, we can identify three regimes: appearance of an initial wavelength, coarsening of the pattern and finally saturation of the ripples. We show that both initial and final wavelengths, as well as the propagative speed of the ripples, are linear functions of the wind velocity. Investigating the evolution of an initially corrugated bed, we exhibit non-linear stable solutions for a finite range of wavelengths, which demonstrates the existence of a saturation in amplitude. These results contradict most of the models.
\end{abstract}


\maketitle

The surface of aeolian sand dunes is not smooth but is usually formed into regular patterns of ripples, transverse to the wind \cite{B41}. Their wavelength ranges from the centimeter to the meter with a constant aspect ratio ($\simeq4\%$) \cite{EEW75}. Although many different models have been proposed to explain the formation and evolution of aeolian ripples \cite{A87,A90,NO93,TCB98,HM99,P99,CMRV00,KBH00,YBP04}, few field observations \cite{B41,EEW75,S63} and controlled experiments \cite{SL78,A90,WG93} have been performed so far. By contrast with  subaqueous dunes or ripples which result from a hydrodynamic instability induced by the interaction between shape and flow \cite{B41}, aeolian ripples are of  different nature and result from a screening instability. When the `saltons' -- high energy grains -- collide the bed, they eject grains of smaller energy, `reptons'. The windward slope of a small bump is submitted to more impacts than the lee slope, so that the flux of reptons is higher uphill than downhill and makes the bump amplify.

Most of the models agree for the linear stage of the instability -- see \cite{CMRV00} for a pedagogical review. They assume that the reptons remain trapped by the bed after a single hop of length $a$, distributed according to a distribution $P(a)$. The saltons are  considered as an external reservoir which brings energy into the system. On this basis, the most unstable wavelength $\lambda_0$ is found to scale on the reptation length $\bar{a} = \int \! da \, a \, P(a)$. Besides, the most recent investigations of sand transport, both experimental \cite{IR99,RVB00} and theoretical \cite{AH88,A04}, indicate that $P(a)$ is independent of the wind shear velocity $u^*$, which implies that $\bar{a}$ scales on the grain diameter $d$. The initial wavelength of the ripples $\lambda_0$ is thus expected to be independent of the wind strength.

\begin{figure}[t!]
\begin{center}
\includegraphics{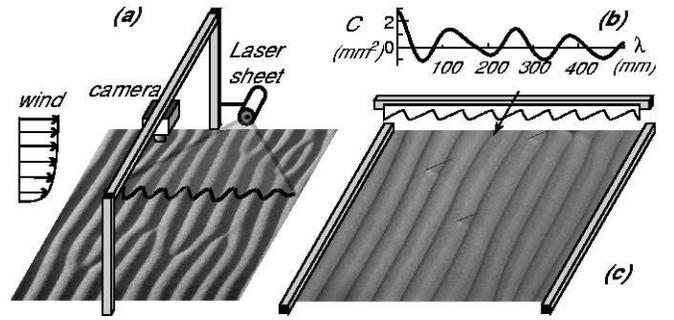}
\end{center}
\caption{Experimental set up both in the field and in the wind tunnel. ({\bf a})  The profile $h(x)$ is given by the inclined laser sheet. ({\bf b}) Autocorrelation function $C(\lambda,t) = <h(x,t) h(x+\lambda,t)> - <h(x,t)>^2$ of the profile. The position of the first peak gives the mean wavelength $\lambda$. The amplitude is defined by $A(t)=2\sqrt{2C(0,t)}$. ({\bf c}) Method for engraving periodic profiles (the lateral bars are removed once the pattern is engraved).}
\label{setup}
\end{figure}

In the non-linear regime, aeolian ripples have been one of the most extensively studied systems presenting coarsening. Two scenarii are evoked. The wavelength can increase by a phase instability (negative diffusion) leading to a continuous stretching of the pattern \cite{PM04}. Alternatively, ripples may be considered as locally interacting objects that undergo successive merging \cite{WG93,AAKESU02}. In both cases, one predicts a perpetual increase of  $\lambda(t)$ as the logarithm of time or as a power law of small exponent. Such a logarithmic behaviour has been argued to be compatible with some wind tunnel data \cite{WG93}. However, common field observations clearly evidence steady patterns driven by a wind permanently blowing for hours. `Fully developped' ripples are also reported in experiments of \cite{SL78}, but not studied in details. Some possible reasons to explain the discrepancies between models and field observations are discussed in \cite{A90}, and to the best of our knowledge, only one recent model predicts a saturation of ripples wavelength at long time \cite{YBP04}. The goal of this paper is to investigate experimentally the precise conditions of saturation, and to study the properties of the saturated states.

\begin{figure*}[t!]
\begin{center}
\includegraphics{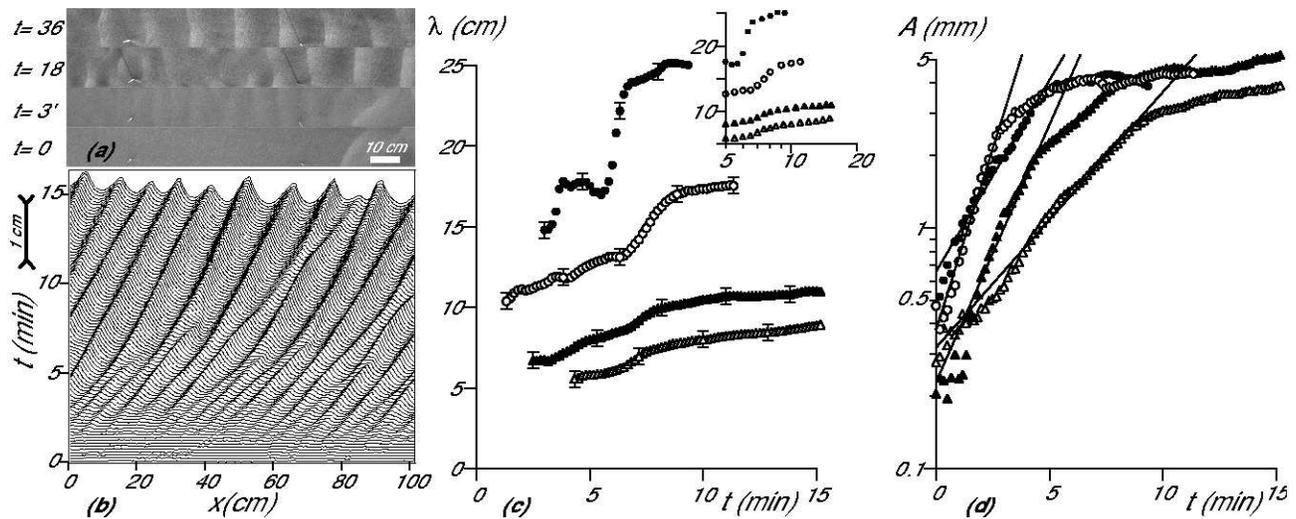}
\end{center}
\caption{Evolution of an initially flat bed. ({\bf a}) Pictures at different times for field experiments $u^* = 1.3~u_{th}$, ({\bf b}) Spatio temporal diagram in the wind tunnel for $u^* = 1.4~u_{th}$. Time evolution of ({\bf c}) wavelength $\lambda(t)$ and ({\bf d}) amplitude $A(t)$ from a flat bed for different values of the shear velocity: $u^* = 1.3~u_{th}$ ($\triangle$), $u^* = 1.4~u_{th}$ ($\blacktriangle$), $u^* = 1.8~u_{th}$ ({\Large $\circ$}), $u^* = 2.3~u_{th}$ ({\Large $\bullet$}). Inset: time log scale.}
\label{spatiolibre}
\end{figure*}

\emph{Experimental set-up} -- We have performed experiments both in the wind tunnel of the \textsc{cemagref} in Grenoble ($1$~m wide, $0.5$~m high and $4.5$~m long) and in the barchan field extending between Tarfaya and Laayoune in Morocco. In the wind tunnel, the air flow is well controlled and can reach $u^* \sim 0.7$~m/s but the sand flux is not fully saturated due to the limited length of the sand bed ($3$~m). In the field, the sand supply is unlimited and the flux is saturated but the wind strength fluctuates. The grains we used in the lab -- so-called Hostun sand --
are $120 \pm 40~\mu$m in diameter and are rather angular as this sand comes from a quarry. The grains in the dune field  is more round and smooth, made of quartz/lime mixture, and with $d\sim 180 \pm 35~\mu$m. In both situations the measured shear velocity threshold for saltation is $u_{th} \sim 0.22$~m/s. To measure the sand surface deformation, the same method is used both in the field and in the wind tunnel. A laser sheet is inclined at some low angle to the bed (Fig.~\ref{setup}a) and pictures are taken from above using a $2240\times1680$ digital camera. The profile $h(x,t)$ is detected by a correlation method which insures a precision of $40~\mu$m. The patterns obtained are usually quasi-periodic and can be characterized by an average wavelength $\lambda$ and an amplitude $A$ (Fig.~\ref{setup}b).

\begin{figure*}[t!]
\includegraphics{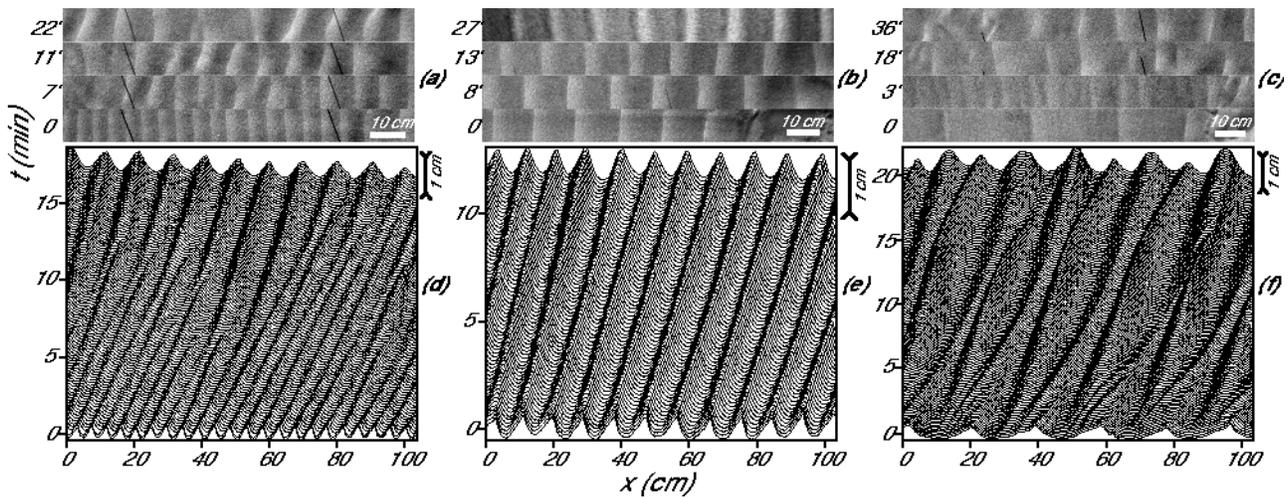}
\caption{Evolution  of an initially corrugated bed. Pictures from the field for $u^* = 1.3~u_{th}$ and initial wavelength ({\bf  a}) $\lambda=5$~cm; ({\bf  b}) $\lambda=13.5$~cm; ({\bf  c}) $\lambda=30$~cm. Spatio temporal diagram in the wind tunnel for $u^* = 1.3~u_{th}$ and ({\bf  d}) $\lambda=5$~cm; ({\bf  e})  $\lambda=9.3$~cm; ({\bf  f}) $\lambda=19.5$~cm.}
\label{spatioforcestunnel}
\end{figure*}

\emph{Instability of a flat bed} -- We first study the time evolution of an initially flat bed. The dynamics in the field is shown in Fig.~\ref{spatiolibre}a and the spatio-temporal diagram obtained in the wind tunnel is displayed in Fig.~\ref{spatiolibre}b. In both cases, the initially flat surface deforms and coarsening is observed. The time evolution of wavelength and  amplitude for experiments in  the wind tunnel are plotted in Figs.~\ref{spatiolibre}c,d. Identifying a linear regime is not an easy task. A gentle  linear instability  is usually characterised by an exponential growth of the amplitude and a plateau in wavelength. Here, tiny structures with a small wavelength show up after a short time ($\sim 1$~min for this run). Their growth in amplitude is consistent with an exponential when $A$ is smaller than, say, $1$~mm i.e. $6$ grain sizes. However, a peak clearly appears in the autocorrelation function  $C(\lambda,t)$ only close to the end of this period. This tiny plateau still allows to define the initial wavelength $\lambda_0$. As soon as they are visible, the ripples start merging so that the pattern exhibits coarsening. The increase of amplitude and wavelength progressively slows down and after some time (typically $\sim 10$~min), the pattern tends to `saturate': the ripples essentially propagate without changing shape and amplitude anymore. We take $\lambda_\infty$ as the average of $\lambda$ over the last two minutes. As emphasized by the log scale in time, this ultimate regime could also be interpreted as a slow drift in $\lambda$ \cite{WG93}. Settling this issue would require at least one more decade in time (few hours), which is difficult to achieve in our wind tunnel due to the global erosion of the bed (a sand layer few centimetres thick disappears after typically half an hour) and to the convective nature of the instability (the entrance conditions can influence the measurement zone).
\begin{figure}[b!]
\includegraphics{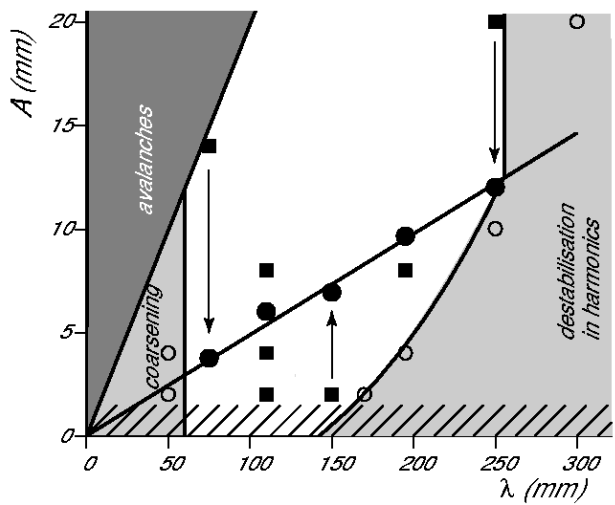}
\caption{Stability diagram of ripples of initial wavelength $\lambda$ and amplitude $A$ for $u^*=1.3~u_{th}$. Outside the locking zone ({\Large $\circ$}), the pattern destabilizes by generation of harmonics (right) or sub-harmonics (left). Starting in the central region ($\blacksquare$), $\lambda$ stays constant and $A$ readapts to reach a stable steady solution ({\Large $\bullet$}). Very large amplitudes cannot be imposed as avalanche processes take place. Hatched region is beyond experimental access.}
\label{ALambda}
\end{figure}
%

\emph{Evolution of an initially corrugated bed} -- To bypass the problem, we have performed experiments in which we do not start from a flat bed but from a corrugated surface. A periodic ripple profile is initially engraved with a cardboard pattern on the bed and is then let free to evolve (Fig.~\ref{setup}c). In the wind tunnel, we have been able to systematically study the influence of the initial wavelength and amplitude, the shape remaining that of a developed ripple. The first result concerns the case when the engraved pattern has the same  wavelength $\lambda_\infty$ and same amplitude $A_\infty$ as measured in `natural conditions', i.e. starting from a flat bed. The ripples do not evolve at all and the pattern purely propagates (Fig.~\ref{spatioforcestunnel}e), showing that a saturated state exists. If we start from the same $\lambda$ but with a larger or a smaller amplitude, the pattern converges back to the same shape ($A_\infty$, $\lambda_\infty$) by adjusting its amplitude. If we start from the same $A$ but a  slightly different $\lambda$, the patterns keeps this new wavelength and selects a new amplitude (Fig.~\ref{ALambda}). We can thus conclude that, depending on the initial conditions, the pattern converges toward different stable non-linear solutions. The saturated state observed starting from a flat bed is one solution among the finite range of wavelengths for which stable ripple patterns can be observed. If we start from a too  large $\lambda$, the pattern destabilizes by changing its wavelength, forming  substructures on the windward slope (generation of harmonics) as shown in Fig.~\ref{spatioforcestunnel}f. Too small $\lambda$ are also unstable as the pattern coarsens due to merging (period doubling) (Fig.~\ref{spatioforcestunnel}d). In the ($A$, $\lambda$) diagram representing the different initial conditions (Fig.~\ref{ALambda}), one can then determine a stable region  where the engraved pattern does not change wavelength, an unstable region leading to coarsening for $\lambda \lesssim 60$~mm and an unstable region leading to a windward slope instability. These observations in the wind tunnel hold for our field experiments: Figs~\ref{spatioforcestunnel}a,b,c show that the engraved ripples follow the same spatio temporal evolution as in the wind tunnel when the wavelength is smaller, close to or larger than the `natural' one. Due to wind fluctuations we have not been able to systematically study the influence of the initial amplitude.
\begin{figure*}[t!]
\includegraphics{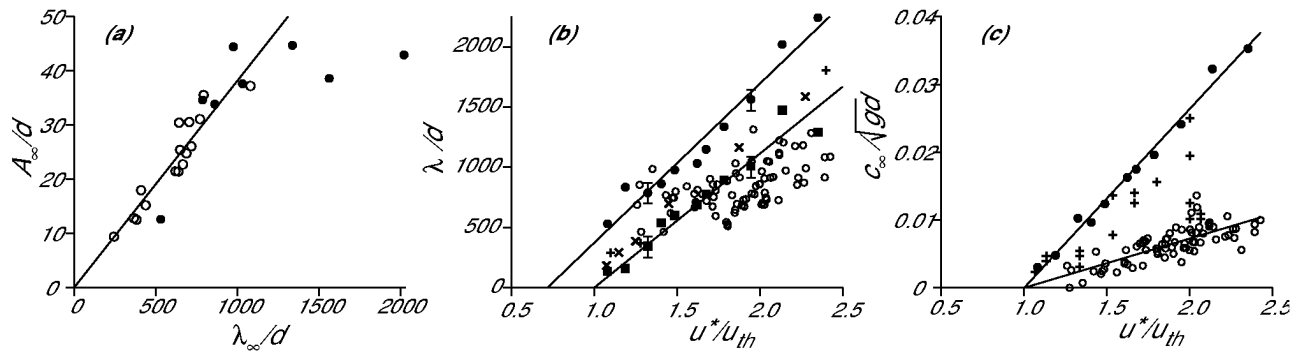}
\caption{({\bf a}) $A$ as a function of $\lambda$ for different wind velocities in the wind tunnel ({\Large $\bullet$}) and in the field ({\Large $\circ$}). ({\bf b}) Initial ($\blacksquare$) and final ({\Large $\bullet$}) wavelength  (wind tunnel data) and field final wavelength ({\Large $\circ$}) as a function of the imposed flow velocity $u_*$. ({\bf c}) Propagation velocity of developped ripples as a function of the wind speed for free evolving ripples in the wind tunnel ({\Large $\bullet$}) and in the field ({\Large $\circ$}). Data represented by {\bf +} (resp. $\times$) come from \cite{S63} (\cite{SL78}), for grains of size $d \sim 310~\mu$m ($150~\mu$m).}
\label{machins_deu}
\end{figure*}

\emph{Parametric study} -- Once we have shown that the final pattern observed starting from a flat bed corresponds to a saturated state, we can systematically study the influence of the wind on the natural ripples characteristics. Fig.~\ref{machins_deu}a shows the selection of the aspect ratio for different wind strengths. While the aspect ratio remains constant in the field ($A\sim0.04\lambda$), the amplitude saturates in the wind tunnel at large $u^*$. This discrepancy may be related to the unsaturated state of the sand flux. The initial and final wavelengths scaled by the particles diameter are plotted as a function of the wind speed in Fig.~\ref{machins_deu}b. By contrast with  standard predictions, they both increase linearly with $u^*$. This linear behaviour was also found by \cite{SL78}. Interestingly, the initial wavelength vanishes at the transport threshold and both $\lambda_0(u^*)$ and $\lambda_\infty(u^*)$ are parallel. The same tendency holds for field measurements, although more scattered. In a similar way, the rescaled phase velocity $c_\infty/\sqrt{gd}$ of saturated ripples increases linearly with the wind (Fig.~\ref{machins_deu}c). Our data then suggest that $\lambda_0/d  \propto (u^*/u_{th} -1)$ and  $c_\infty/\sqrt{gd}\propto (u^*/u_{th} -1)$. In order to test the scaling with $d$ which simply comes from dimensional analysis, we have also plotted in Figs.~\ref{machins_deu}b,c data coming from \cite{S63} and \cite{SL78}. The  collapse is good, showing that $d$ is the relevant length scale. The prefactor in these linear relations certainly depends on the density ratio $\rho_{\rm sand}/\rho_{\rm air}$, which however is difficult to vary. Comparison with martian ripples formed in an atmosphere $80$ times lighter \cite{Setal05} could be of interest to get deeper insight in the scaling.

\emph {Conclusion} --  Two main conclusions can be drawn from this study. First, the linear dependence of the most unstable wavelength with the wind speed rules out our theoretical understanding of ripples formation and sand transport in general. As $\lambda_0$ turns out to be much larger than the value predicted by Anderson-like models ($6\bar a\sim60d \sim 10$~mm), there should exist a yet unknown mechanism stabilizing small wavelengths. Second, we have shown the existence of a whole family of stable non linear solutions. Future models should then be able to recover the range of wavelengths for which ripple patterns are stable, as well as the saturation of $\lambda$ when starting from a flat bed.

In this work, we have studied the one dimensional dynamics of the ripples along the wind direction. However, a complex  dynamics exists in the transverse direction too and the formation of defects is known to play a significant role in coarsening processes  and in the route towards regular pattern \cite{WG93,AAKESU02}. It then suggests to investigate the density of defects to characterize the degree of saturation of the pattern. Finally, ripples could be of practical interest for field experimentalist as a non intrusive measurement of the reptation sand flux.

We wish to thank  Florence Naaim and the ETNA group of the \textsc{cemagref} in Grenoble for the use of their wind tunnel and for their kind help during the experiments. This study was supported by the french ministry of research through an `ACI Jeunes C'.


\end{document}